# Two Permanently Congruent Rods May Have Different Proper Lengths

Moses Fayngold

*Department of Physics, New Jersey Institute of Technology, Newark, NJ 07102*

We scrutinize congruence as one of the basic definitions of equality in geometry and pit it against physics of Special Relativity. We show that two non-rigid rods permanently kept congruent during their common expansion or compression may have different instantaneous proper lengths (when measured at the same time of their respective reference clocks) if they have different mass distributions over their lengths. Alternatively, their proper lengths can come out equal only when measured at different but strictly correlated moments of time of their respective clocks. The derived expression for the ratio of instantaneous proper lengths of two permanently congruent changing objects explicitly contains information about the objects' mass distribution. The same is true for the ratio of readings of the two reference clocks, for which the instantaneous measurements of respective proper lengths produce the same result. In either case the characteristics usually considered as purely kinematic depend on mass distribution, which is a dynamic property. This is a spectacular demonstration of dynamic aspect of geometry already in the framework of Special Relativity.

The concept of size, and more specifically, proper length, is usually applied to objects whose size is either fixed or changing slowly enough to be considered as constant within a certain time interval. Within these restrictions, establishing the lasting congruence of two objects automatically means their complete geometrical identity during the corresponding time interval. This remains true for rapidly-changing objects, if we still consider them in the framework of non-relativistic mechanics. However, in Relativity, the congruence of two objects does not by itself guarantee their complete geometrical identity. More accurately, the two objects may be geometrically identical but still have their proper lengths not equal to each other when measured at the same moment of their respective times, which enables one to say that their geometrical identity is not complete. This peculiar feature is due to the fact that in relativity one has to distinguish between the size of an object as observed from an arbitrary reference frame and its *proper* size. In the following discussion we restrict ourselves to only one dimension and will accordingly consider congruent straight rods.

As a first step, we note that permanent congruence of two *changing* objects A and B considered here is physically different from the congruence of two disks in different states of rotation, as discussed in the Ehrenfest paradox [1-8]. The former requires permanent coincidence of any small adjacent parts $\delta A$ and $\delta B$ at least on the periphery of the congruent objects (local congruence), which implies the absence of relative motion between these parts. The latter constitutes only global congruence while any (except those at the center) instantly adjacent small parts $\delta A$ and $\delta B$ of two disks A and B are in relative motion with respect to one another.

Next, we apply the definition of proper length to rods of varying size. In this more general case, the proper length is a function of time, and its measurement must meet an additional requirement. For any rod, the proper length is the distance between its end points measured in its rest frame. If the rod is rigid and we are in its rest frame, the requirement that the end points should be marked simultaneously is redundant (as opposed to measuring the Lorentz-contracted length of a moving rigid rod). However, for a non-rigid rod with changing size the requirement of simultaneous markings becomes absolutely imperative even in its rest frame. For an object with moving parts, such frame is defined as the reference frame in which the object's net momentum is zero.

We now consider the implications of this generalized definition. To avoid complications associated with the elastic energy [7-9] consider a simplified model of the rod as a linear array of non-interacting point masses. For our purposes, the requirement of local congruence can be retained only for the edge masses of the respective rods. Actually, a rod represented by two masses at its edges will be sufficient to illustrate the main point.

Consider two such rods A and B, with their respective edges ($A_1$, $B_1$) at one end and ($A_2$, $B_2$) at the other being permanently coincident (Fig. 1). If the mass distribution is the same for both rods, no problems arise. Both rods, while having their size changing, remain at each moment identical physically and geometrically. Their proper lengths can be represented by the same common function of time:

$$l_A^0(t) = l_B^0(t) = l^0(t) \tag{1}$$

The situation becomes more subtle when the corresponding mass distributions of the rods are different. Suppose rod A is uniform, while rod B is not. Accordingly, rod A can be represented by 2 equal point masses $m_0$ at its edges $A_1$ and $A_2$; rod B is represented by two unequal point masses: $m_0$ at $B_1$ and $M_0$ at $B_2$. In all cases the subscript "0" indicates the rest mass.

Let both rods expand while remaining congruent (Fig. 2). Consider the whole process from the frame A[1]. In this case the center of mass of rod A remains exactly at its middle point O, which we can take as the origin. The center of mass O' of B is, first, closer to the more massive end $B_2$ and secondly, is moving relative to O.

Since the length of an expanding rod is a function of time, we have to deal with its instantaneous length. Due to the permanent congruence of the two rods, their instantaneous lengths are equal at any time in any reference frame. This, however, is *not* true regarding their *proper* lengths. We can show this by applying the basic definition of proper length to each rod in its respective RF. In frame A, the length of rod A is its proper length; the length of rod B (when also measured in frame A), however, is not, since this rod is moving as a whole relative to O. Therefore measurement of both rods' length in system A does not constitute the *proper* length measurement for rod B. Reversing the argument, we can say that the instantaneous length measurement in system B for both rods constitutes the *proper* length measurement for B but not for A.

For the most simple quantitative description, it is sufficient to consider the expansion of both rods starting from the origin O (that is, both O and O' are coincident at the initial moment $t = t' = 0$) and continuing at a constant rate, so that in frame A the edges $A_1$ and $A_2$ are both receding from their origin O at the same speed $v$ (the same holds in this frame for $B_1$ and $B_2$ due to the permanent coincidence of the rods). In the language of Astrophysics, this can be considered as a simplified geometrical model of birth of two parallel one-dimensional universes in one common Big Bang, and their subsequent common expansion[2].

In frame A, the instantaneous proper length of A is determined as

$$l_A^0(t) = 2vt \qquad (2)$$

The length of rod B measured in this frame is

$$l_B(t) = 2vt = l_A^0(t) \neq l_B^0(t) \qquad (3)$$

The left side of these equations reflects the congruence of the rods. However, as mentioned above, the length $l_B(t)$ of rod B is *not* its proper length $l_B^0(t)$ since the frame A is not its rest frame. According to the generalized definition of instantaneous proper length, $l_B^0(t)$ under the assumed conditions can be measured directly in frame B as

$$l_B^0(t') = (v_1' + v_2')t' \qquad (4)$$

Here $v_1'$ and $v_2'$ are the speeds in this frame of the edges $B_1$ and $B_2$, respectively; $t'$ is time in this frame. According to the transformation rule for velocities,

---

[1] From now on, "Frame A" will stand for the rest frame of rod A, and similarly for B.
[2] Both universes here have their centers and expand in the external space, not together with space.

$$v_1' = \frac{v + V_C}{1 + \frac{vV_C}{c^2}} \; ; \qquad v_2' = \frac{v - V_C}{1 - \frac{vV_C}{c^2}} , \qquad (5)$$

where $V_C$ is the relative velocity of systems A and B, that is, the velocity of center of mass of rod B relative to A. For a system of non-interacting particles,

$$\mathbf{V}_C = \frac{\mathbf{P}}{M} = \frac{\sum_j m_j \mathbf{v}_j}{\sum_j m_j} \qquad (6)$$

where $\mathbf{P}$ is the net momentum of the system, $M$ is its relativistic mass, $v_j$ and $m_j = m_{0j}\gamma(v_j)$ are, respectively, the velocity and relativistic mass of $j$-th particle, and $\gamma(v_j)$ is the corresponding Lorentz factor [10]. In our simple model of an expanding rod represented by two end masses $m_{01} = m_0$ and $m_{02} = M_0$ flying apart with equal speeds $v_1 = v_2 = v$ (in frame A), this yields

$$V_C = \frac{P_B}{M_B} = \frac{\mu - 1}{\mu + 1} v , \qquad \mu \equiv \frac{M_0}{m_0} \qquad (7)$$

Combining Equations (4) and (5), we obtain

$$l_B^0(t') = 2vt' \frac{1 - \beta_c^2}{1 - \beta^2 \beta_c^2} , \qquad \beta \equiv \frac{v}{c} , \qquad \beta_c \equiv \frac{V_C}{c} \qquad (8)$$

Suppose we have a reference clock at the center of mass of rod A, and an identical clock at the center of mass of rod B. By the initial condition, both clocks are set to read the zero time at the onset of the expansion, when they are instantly coincident. In view of the equivalence of all inertial RF, it is tempting to think that two observers in the rest frames of the respective congruent rods would obtain the same result if they measure the rods' proper lengths when their respective clocks read the same time. This would be the case in non-relativistic mechanics, according to which time is absolute and dynamical characteristics of a system do not affect the geometry; in particular, the congruence of two rods means their geometrical identity irrespective of any peculiarities in their mass distribution.

The situation is quite different in the relativistic domain. Setting $t = t'$ in the equations (2), (8), we obtain

$$l_B^0(t) = l_A^0(t) \frac{1 - \beta_c^2}{1 - \beta^2 \beta_c^2} \equiv \frac{l_A^0(t)}{\gamma^2(\beta_c)(1 - \beta^2 \beta_c^2)} \qquad (9)$$

We see that $l_B^0(t) \neq l_A^0(t)$, that is, the Eq. (1) does not hold in this case. The two proper lengths turn out to be described by *different functions of time.*

We can also use (7) to recast the Eq. (9) in the form:

$$l_B^0(t) = \frac{1 - \left(\frac{\mu-1}{\mu+1}\right)^2 \beta^2}{1 - \left(\frac{\mu-1}{\mu+1}\right)^2 \beta^4} \, l_A^0(t) \tag{10}$$

In this form, the ratio $\xi$ of the instantaneous proper lengths of the two permanently congruent rods measured at the same moment of their respective times is an explicit function of their mass distribution: $\xi = \xi(\mu, \beta)$. As seen from (10), this function

$$\xi(\mu,\beta) \equiv \frac{l_B^0(t)}{l_A^0(t)} = \frac{1 - \left(\frac{\mu-1}{\mu+1}\right)^2 \beta^2}{1 - \left(\frac{\mu-1}{\mu+1}\right)^2 \beta^4} \leq 1 \tag{11}$$

is never greater than 1 (Fig. 3). Its upper extreme 1 is only reached in the limit when $\beta \to 0$ (the rods stop expanding) or when $\mu \to 1$ (the rods, apart from being congruent, have the same distribution of mass). But if the rods have different mass distribution and change in size concurrently at a finite rate, then $\xi < 1$, that is, they have different proper lengths. The inhomogeneous rod turns out to be intrinsically shorter than the homogeneous one congruent with it. For instance, if $\mu = 100$ (the right edge of B is 100 times more massive than its left edge) and $\beta = 0.6$, the Eq. (11) gives

$$\xi \equiv l_B^0(t)/l_A^0(t) \approx 0.75 \tag{12}$$

At $\mu \to \infty$ and $\beta \to 1$ (e.g. two identical neutrinos flying apart in A and a similar neutrino at $B_1$ and an atom at $B_2$ in B) we have

$$\xi \to \frac{1 - \beta^2 + 4\mu^{-1}\beta^2}{1 - \beta^4 + 4\mu^{-1}\beta^4} \to \frac{1}{1+\beta^2} \to \frac{1}{2} \tag{13}$$

The limit (13) forms a lower extreme of $\xi$.

Summarizing this part, we see that the ratio of the instantaneous proper lengths of the two rods measured at the same moment of their respective times after the beginning of their expansion can vary within a range

$$\frac{1}{2} \leq \xi \leq 1. \tag{14}$$

As an alternative way of treatment, we can derive the expressions for the proper lengths of the two rods in terms of $v_1'$ and $v_2'$ only. In other words, we can use variables $v_1'$, $v_2'$ instead of $v, \mu$.

First, the Eq. (9) can be written in a slightly different way (in terms of $v_{1,2}'$ and $v$ rather than $V_c$ and $v$) if we use a well-known identity (see, e.g., [8])

$$\gamma(v'_{1,2}) = \gamma(v)\,\gamma(V_C)\left(1 \pm \frac{vV_C}{c^2}\right), \tag{15}$$

where "+" and "–" relate to $v'_1$ or $v'_2$, respectively. Applying this to (10), we obtain

$$l^0_B(t) = \frac{\gamma^2(v)}{\gamma(v'_1)\,\gamma(v'_2)}\,l^0_A(t). \tag{16}$$

Next, we need to express $v$ in terms of $v'_1$, $v'_2$ as well. We have from (5), using notations (8):

$$\beta = \frac{\beta'_1 - \beta_c}{1 - \beta'_1\beta_c} = \frac{\beta'_2 - \beta_c}{1 - \beta'_2\beta_c} \tag{17}$$

Solving this for $\beta_c$ and applying the selection rule $\beta_c \to 0$ at $\beta'_1 \to \beta'_2$ gives

$$\beta_c = \frac{\gamma(\beta'_1)\,\gamma(\beta'_2)\,(1 - \beta'_1\beta'_2) - 1}{\gamma(\beta'_1)\,\gamma(\beta'_2)\,(\beta'_1 - \beta'_2)}, \tag{18}$$

or, in view of identity (15)

$$\beta_c = \frac{\gamma(\beta'_{1,2}) - 1}{\gamma(\beta'_1)\,\gamma(\beta'_2)\,(\beta'_1 - \beta'_2)}, \quad \beta'_{1,2} \equiv \frac{\beta'_1 - \beta'_2}{1 - \beta'_1\beta'_2}, \tag{19}$$

where $\beta'_{1,2}$ is the relative velocity of the edges of the rod B. The sought-for expression for $\beta(\beta'_1, \beta'_2)$ is obtained by putting the solution (19) back into (17) with $\beta$ on the left retained.

An alternative (and more simple) way to find $\beta_c$ in terms of $\beta'_1$, $\beta'_2$ is to consider from the very beginning the motion of rod A in frame B. According to (6), and in view of congruence, the velocity of the center of mass of A in this frame can be immediately written as

$$\mathbf{V}_c = \frac{m_0\,\gamma(v'_1)\,\mathbf{v}'_1 + m_0\,\gamma(v'_2)\,\mathbf{v}'_2}{m_0\,\gamma(v'_1) + m_0\,\gamma(v'_2)}, \tag{20}$$

or

$$\beta_c = \frac{\beta'_1\gamma(\beta'_1) - \beta'_2\gamma(\beta'_2)}{\gamma(\beta'_1) + \gamma(\beta'_2)} \tag{21}$$

As a by-product of our calculations, we notice that if both – (18) and (21) – are correct, then there must be an identity:

$$\frac{\gamma(\beta'_1)\,\gamma(\beta'_2)\,(1 - \beta'_1\beta'_2)}{\gamma(\beta'_1)\,\gamma(\beta'_2)\,(\beta'_1 - \beta'_2)} \equiv \frac{\beta'_1\gamma(\beta'_1) - \beta'_2\gamma(\beta'_2)}{\gamma(\beta'_1) + \gamma(\beta'_2)} \tag{22}$$

One can check that this is, indeed, the case.

Putting the found expression for $\beta_c$ back into (17) yields after some algebra

$$\beta_1 = \beta_2 = \beta = \frac{\gamma(\beta_1')\,\gamma(\beta_2')\,(\beta_1'+\beta_2')}{1+\gamma(\beta_1')\,\gamma(\beta_2')\,(1+\beta_1'\beta_2')}\,. \tag{23}$$

Therefore

$$\beta_1 + \beta_2 = 2\beta = \frac{\gamma(\beta_1')\,\gamma(\beta_2')}{1+\gamma(\beta_1')\gamma(\beta_2')\,(1+\beta_1'\beta_2')}\,(\beta_1'+\beta_2'), \tag{24}$$

and

$$\xi(\beta_1', \beta_2') \equiv \frac{l_B(t)}{l_A(t)} = \frac{\beta_1'+\beta_2'}{\beta_1+\beta_2} = \frac{1+\gamma(\beta_1')\gamma(\beta_2')\,(1+\beta_1'\beta_2')}{2\gamma(\beta_1')\,\gamma(\beta_2')} \tag{25}$$

This is the same result (11), but now expressed in terms of variables $\beta_1'$, $\beta_2'$ (Fig.4). As in (11), the ratio $\xi$ may vary within the range (14). Its naively expected value $\xi = 1$ is only a *special* case attained in the limit $\beta_1' = \beta_2' \equiv \beta' = \beta$. Since in frame B we have

$$m_0\,\beta_1'\,\gamma(\beta_1') = M_0\,\beta_2'\,\gamma(\beta_2')\,, \quad \text{that is,} \quad \frac{\beta_1'\,\gamma(\beta_1')}{\beta_2'\,\gamma(\beta_2')} = \frac{M_0}{m_0} \equiv \mu\,, \tag{26}$$

this limit corresponds to $\mu = 1$, that is, to the case when the rods are physically identical (have the same mass distribution). In other words, the complete geometrical identity for the changing rods in our case reflects their physical identity. Note, however, that physical identity in this case is not required to be complete; the essential requirement is that the ratio of the end masses of rod A must be the same as that of rod B; the total rest masses of the rods may be different. More generally, in order for the changing congruent rods A and B, possibly with different mass distributions and even containing different numbers of point masses, to have equal proper lengths at equal moments of their respective times, there must exist a RF in which

$$\sum_{i=1}^{N_A} m_i^{(A)}\,\mathbf{v}_i^{(A)}(t) = \sum_{j=1}^{N_B} m_j^{(B)}\,\mathbf{v}_j^{(B)}(t) = 0 \tag{27}$$

Here again we assume a rod as a linear chain of non-interacting point masses, and use the same notations as in (6). In order for (27) to hold for a non-uniform rod, the corresponding number $N_A$ or $N_B$ must be greater than 2. If the condition (27) is satisfied, then, by definition, the corresponding RF is the common rest frame of either rod; and by congruence (at least peripheral!), the rods have the same proper lengths at any moment of their common proper time *t*.

According to (9), the proper length $l_B^0(t)$ cannot be determined from the instantaneous length $l_B(t) = l_A^0(t)$ (measured in frame A) by using a known expression (Lorentz contraction effect):

$$l_B^o(t) \neq l_B(t)\,\gamma(\beta_c) = l_A^0(t)\,\gamma(\beta_c)\,, \tag{28}$$

because the considered situation is essentially different from the case of two rigid rods moving relative to one another. It can, however, be defined under certain conditions as the invariant (proper distance) $\Delta x^0$ in a space-like 4-interval

$$\left(\Delta x^0\right)^2 = \left(\Delta x\right)^2 - c^2 \Delta t^2 = \left(l_B^0\right)^2, \tag{29}$$

even though the proper distance is generally *not* the same thing as the proper length [8].

It might at first seem unclear how a mere change in mass of one of the rod's edges can physically affect its instantaneous proper length. This becomes immediately clear if we recall that the proper length of an object is one of its characteristics observed in its rest frame. Applying this to rod B and using (6), we see that a change in its mass distribution automatically changes position *and* speed of its center of mass and thereby its rest frame, if the constituting masses are moving. If such a switch to another rest frame occurs due to a change in mass distribution, this immediately affects the instantaneous proper length of an expanding (or contracting) rod, *without any additional shift* of either of its edges.

As a physical model (or a thought experiment) illustrating this point, we may consider a *transition* from a pair of the two rods (A, A) which are initially *totally identical* both geometrically and physically, to a pair of rods (A, B) which remain only geometrically congruent, but become different in their respective mass distributions. Namely, the right edge of one of the rods instantaneously becomes more massive due to the acquisition of an additional rest mass. Such an acquisition can occur, for instance, in the process of absorption of two identical particles oncoming symmetrically toward the right edge of the rod scheduled to become "rod B", with the *x*-components of their velocities both equal to *v*, where *x* is directed along the rod (Fig. 5a). Suppose that both particles are absorbed by the edge $B_2$ of rod B. If $\delta M_0$ is the rest mass of the system of incident particles, their absorption immediately changes the rest mass of $B_2$ from $m_0$ to

$$m_0 \to m_0 + \delta M_0 \equiv M_0, \quad \delta M_0 = 2\delta m_0 \gamma(\tilde{v}'), \tag{30}$$

where $\tilde{v}' = \tilde{v}'_1 = \tilde{v}'_2$ (Fig. 5b). This change can be directly measured by an observer in the auxiliary RF co-moving with $B_2$ (and thereby $A_2$). For this observer, the system of incident particles has the zero net momentum (Fig. 5b), and after being absorbed it does not change either initial momentum or velocity (which are also zero in this frame) of the rod's edge. It does, however, change its rest mass according to (30).

The nature of the absorbed particles is immaterial. They might be either massive particles or high-energy photons or laser pulses. In either case the additional rest mass $\delta M_0$ includes the $\delta \mathcal{E}/c^2$ term, where $\delta \mathcal{E}$ is the kinetic energy of incident particles in the auxiliary frame. In case of photons or laser pulses, this kinetic energy is the only contributor to the rest mass, that is (assuming two monochromatic photons), $\delta \mathcal{E}/c^2 = 2\hbar\omega/c^2 = \delta M_0$. In other words, in this case a system of two massless particles has a non-zero rest mass $\delta M_0$.

In the initial frame A the right edge of rod B instantaneously acquires additional momentum, but this acquisition is *only* due to an increase in its rest mass and thereby rest energy, whereas its velocity does not change. And nevertheless, *without any jump of velocity*, the proper length of rod B changes instantaneously right after the absorption. The change is due to the instant switch from frame A, which originally was also the rest frame of rod B, to its new rest frame after the absorption. Thus, only change of mass distribution alone over the rod may automatically and instantaneously change its proper length. In addition to the arguments presented in [7, 8, 11-15], this gives yet another example of intimate connections between geometry and dynamics within SR.

An extension of the above argument leads to a farther step towards a more complete description of this connection. Namely, we now must consider in more detail the temporal dimension of space-time. We have found that the *complete* congruence of the two rods (equality of their instantaneous proper lengths) cannot be established by performing the appropriate length measurements at the same moment $t' = t$ of proper time of their respective reference clocks. This negative result may appear to contradict the equivalence of all inertial RF, but it does not. All inertial RF are equivalent with respect to the laws of nature, not with respect to numerical values of some physical characteristics that may be relative.

The results (9) – (11) suggest that the congruence of the two rods can be established by measuring their instantaneous proper lengths at *different* moments of their respective times. In other words, for each moment $t$ of proper length measurement in A, there must be a specific moment of time $t' \neq t$ in frame B such that the proper length measurement of rod B at moment $t'$ would give the same outcome. Thus, not only do we have to jump into another RF $A \rightarrow B$ right after the absorption in the above-described thought experiment, but we must also readjust accordingly the schedule of tests in B due to the corresponding change of simultaneity lines; namely, we must find a specific relationship $t' \leftrightarrow t$ for the respective moments of the length measurements in either system, which would produce the same proper length for permanently congruent rods.

Fig. 6 gives the graphical illustration of all these features in the discussed phenomenon. Here $A_1OA_2$ is the world-sheet of rod A. In view of permanent congruence, it is totally overlapped with the world-sheet $B_1OB_2$ of rod B. The cut $A_1(t) - A_2(t)$ represents the instantaneous size (the proper length) of rod A at a moment $t$ of frame A. The cut $B_1(t) - B_2(t)$, which is parallel to $Ox_B$, represents the proper length of rod B at the same moment of B-time. According to (29), this proper length can be found as the proper distance of the interval $(B_1(t), B_2(t))$, and according to (11) or (25), this distance is shorter than $l_A^0(t)$ (it *looks* longer in the figure because the *pseudo-Euclidean* interval is, by necessity, drawn on the Euclidean plane). The cut $B_1(t') - B_2(t')$, on the other hand, represents the rod B at a later moment $t' > t$ of B-time, for which $l_B^0(t') = l_B^0(t)$.

Note that the sought-for relation $t' \leftrightarrow t$, albeit originating from relativity of time, has little if anything to do with the time dilation effect. The latter is represented by the triangle MNO′(t) in Fig. 6. Point N represents the event simultaneous in A with event O′(t) (the reading $t$ by the O′-clock). Since N is above O(t), the event O′(t) happens later than event O(t), and the reference clock A will accordingly read a later moment $t_N = t\,\gamma(\beta_c) > t$. Similarly, the line $M - O'(t)$ (parallel to the spatial axis of B) represents the set of events simultaneous with O′(t) *in B*. The event M, while happening simultaneously with O′(t) in B, is earlier than O(t) (the point M is below O(t) – the reference clock A reads a moment

$t_M = t\gamma^{-1}(\beta_c) < t$). The reading $t$ of the O-clock is the geometric mean of its readings $t_M$ and $t_N$. From the viewpoint of A, the moving O′-clock is ticking slower than stationary O-clock (time dilation); a reciprocal statement can be made by the B-observer. This well-known and extensively discussed *apparent* paradox (see, e.g., [10, 16–18]) reflects relativity of simultaneity between the events separated by a space-like interval. As seen from Fig. 6, the events O($t$) and O′($t$) marked by *the same* readings of the respective local clocks O and O′, are *not simultaneous* from the viewpoint of either frame A or B. In other words, the same readings of the clocks O and O′ in the discussed situation do not mean *simultaneity* of the corresponding events in either of the two frames, just as they do not guarantee equal results in the respective instantaneous measurements of the proper length of the rods.

Now we want to find the moments $t$ and $t'$ of the respective proper times by the above clocks, which may be *not* the same, but which guarantee the same outcome in the instantaneous measurements of the proper length. Since according to (11) (in the considered case of expansion) $l_B^0(t) \leq l_A^0(t)$ for *the same* moment of respective times in both systems, we expect in this case that $t' \geq t$ for *the same* measured proper lengths of both rods.

To find the exact correspondence we return to the original results (2) and (8), but now instead of setting $t' = t$ we require that

$$l_B^0(t') = l_A^0(t), \tag{31}$$

and use this requirement to determine the sought for-relationship $t \leftrightarrow t'$. We then will have the equation

$$\frac{2vt'}{\gamma^2(V_c)(1-\beta^2\beta_c^2)} = 2vt \tag{32}$$

It follows

$$t' = \frac{1-\beta^2\beta_c^2}{1-\beta_c^2} t, \tag{33}$$

or, in view of (7)

$$t' = \frac{1-\left(\frac{\mu-1}{\mu+1}\right)^2 \beta^4}{1-\left(\frac{\mu-1}{\mu+1}\right)^2 \beta^2} t \tag{34}$$

As we had expected, $t'$ is greater than or (in some special cases) equal to $t$. The ratio $t'/t \equiv \eta$ is just reciprocal of $\xi$ and thus ranges in the region

$$1 \leq \eta(\mu,\beta) = \xi^{-1}(\mu,\beta) \leq 2 \tag{35}$$

Also note that relation (34) is, indeed, totally different from the time dilation effect briefly discussed above. And particularly, in contrast with that effect, in which each observer can legitimately claim that it is not his/her, but the moving partner's clock which is ticking slower, in the given case both observers agree that the required length measurement satisfying condition (31) must be carried out at a later moment of the O′-clock in B. In other words, the measurement establishing equal size of two congruent expanding rods turns out to be performed later in time of inhomogeneous rod. In our initial image invoking the two congruent expanding universes, in any experiment establishing their equal proper size, the inhomogeneous universe turns out to be older. In this respect, the original title of this paper can be replaced with "*Two simultaneously born and permanently congruent expanding universes may have different proper age.*"

Thus, even though the world sheets of the congruent rods are identical in space-time, the *cuts* representing their instantaneous proper lengths are generally different. The selection of cuts (and corresponding moments of time at which their proper lengths are the same) is made by Nature according to individual mass distribution within each rod.

*Summary*:

The instantaneous proper lengths of two permanently congruent changing rods turn out to be equal only when measured at generally different but strictly correlated moments of time in their respective rest frames. In the case considered here the B-observer sitting on the *inhomogeneous* rod B has to measure it generally at a later moment of B-time than does the A-observer when measuring rod A.

The ratio of proper lengths measured at the same moment of time of both systems, or the ratio of corresponding times at which these lengths come out equal, are explicit functions of mass distributions in the rods. This indicates the dynamical underpinnings of kinematics.

It is true that the space-time geometry in SR remains Lorentzian in its own right, without any reference to dynamics. However, the most fundamental features that *make* it Lorentzian, – namely, Lorentz-invariance and pseudo-Euclidean form of a space-time interval ("pseudo-Pythagorean" theorem (29)) – originate from specific behavior of temporal and spatial projections of the interval. And, as shown in [7, 8, 11-15] and in the current paper, this behavior turns out to be connected with dynamics. We conclude that in Special Relativity, space-time geometry and dynamics cannot be considered as totally independent characteristics of the world. In particular, such property of an object as proper length turns out to have essentially dynamic "ingredients". Even with interactions totally neglected, the dynamic aspects of geometrical properties of objects are present already in the framework of Special Relativity. Based on the obtained results, one can expect that the presence of dynamics will be even more impressive in case of interactions, when one would need to explicitly introduce into the picture the energy and momentum of both—particles and their fields.

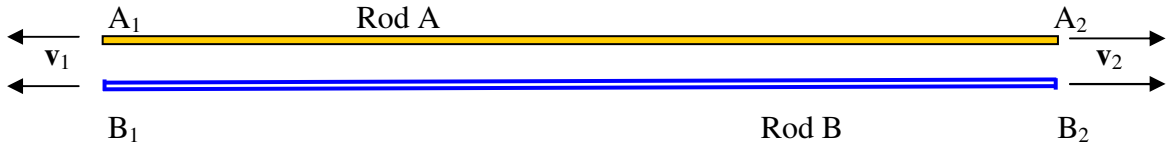

**Fig. 1**
Two congruent and concurrently expanding rods.

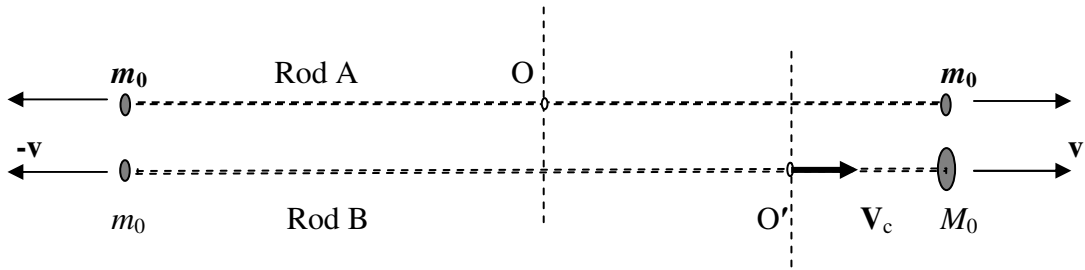

**Fig. 2**
Two congruent and concurrently expanding rods.
The rods are represented by masses concentrated at their edges. Mass distribution of rod B is non-uniform, which is represented by the rest masses at its edges being not equal. The depicted expansion is observed here from the rest frame of rod A. In this frame, the center of mass O′ of rod B, and accordingly, its rest frame, are moving with velocity $V_c$.

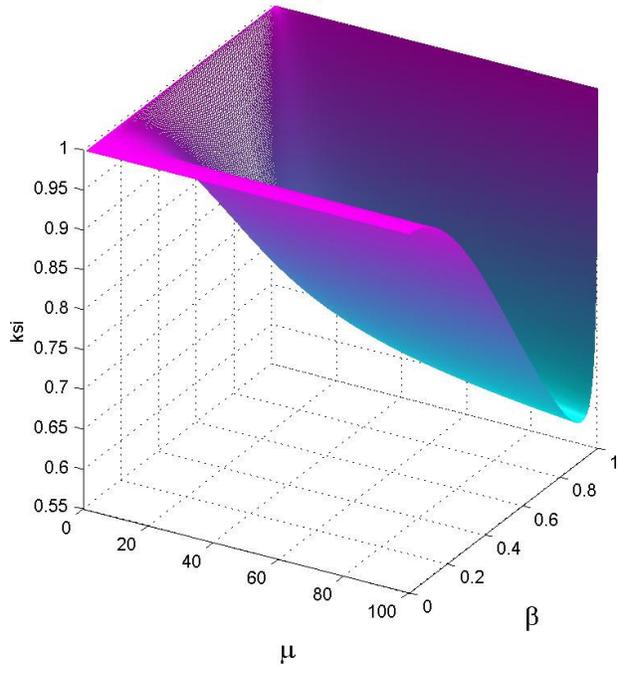

Figure 3

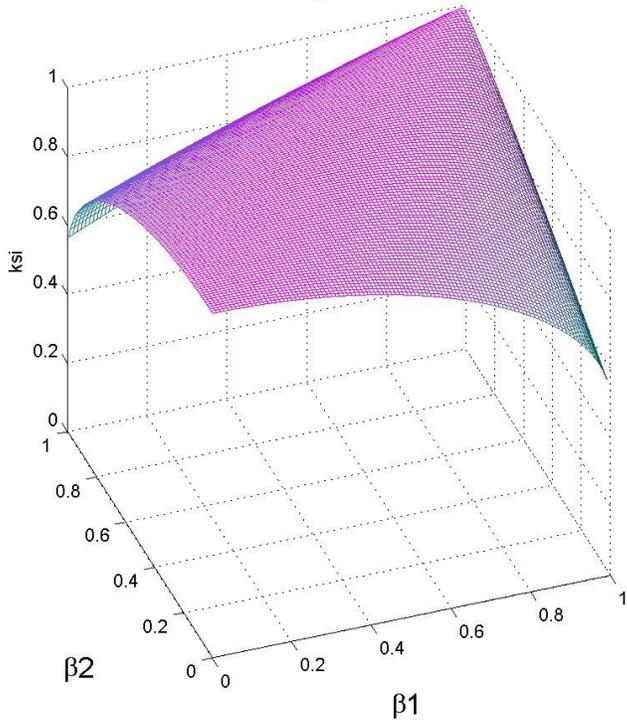

Figure 4

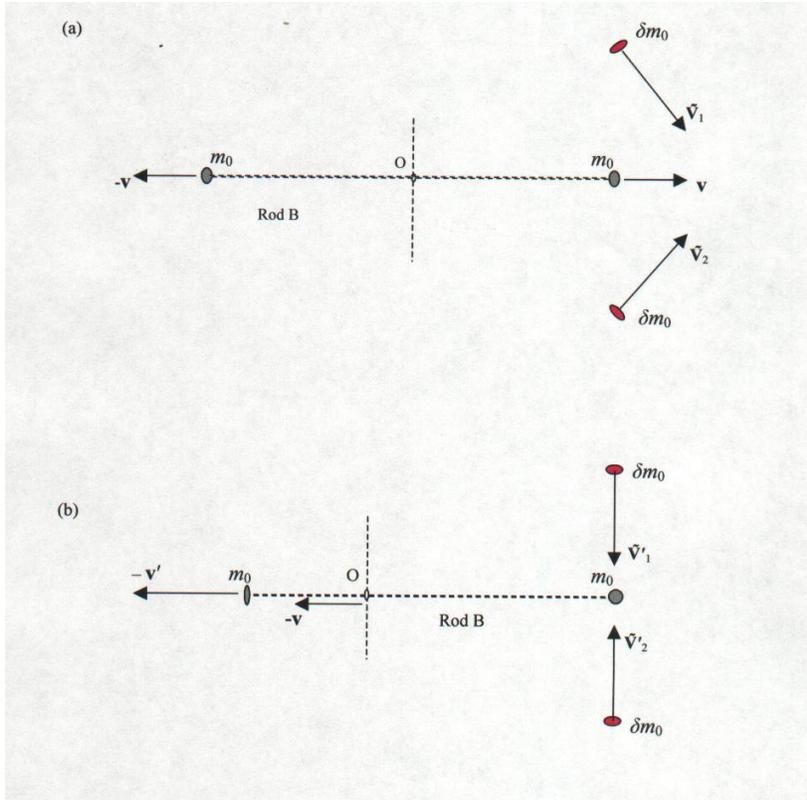

**Fig. 5**

A possible mechanism of rapid change of mass distribution in rod B by absorption of two external particles. The particles' masses and velocities are such that their absorption, while breaking uniformity of the rod, does not affect its expansion rate. It does, however, affect its instantaneous proper length without any additional shift of its edges, by changing the speed of its center of mass and thereby its rest frame.
  (a)  View from the reference frame A (before the absorption).
  (b)  The same pre-absorption stage as viewed from the auxiliary RF co-moving with the right edge of rod A (and B).

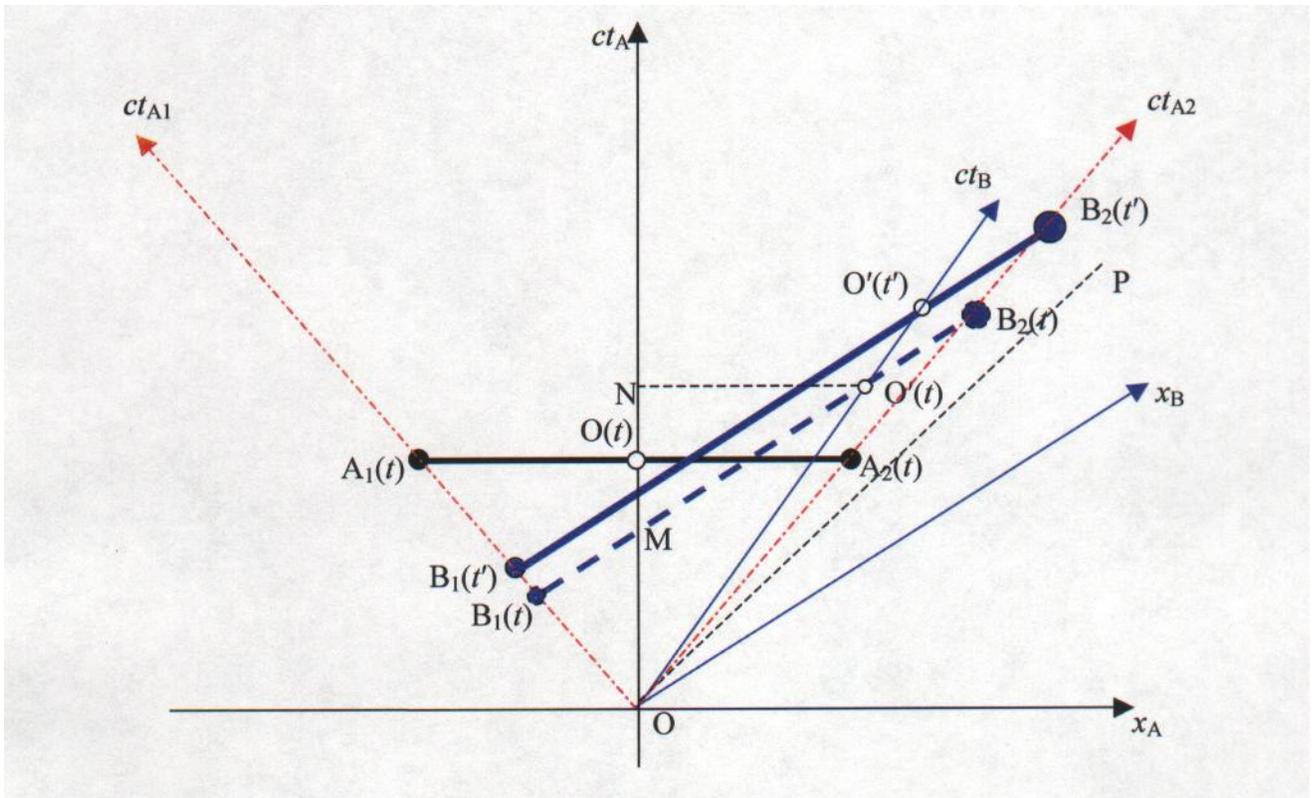

**Fig. 6**
Space-Time diagram of the proper length measurements for two congruent rods.
The diagram (not exactly to scale) represents the case with $\mu = 7$, $\beta = 0.8$.
O-$ct_A$ and O-$x_A$ are respectively the temporal and spatial axes of frame A.
O-$ct_{A1}$ and O-$ct_{A2}$ are the world-lines of the left and right edges, respectively, of both rods.
O-$ct_B$ and O-$x_B$ are respectively the temporal and spatial axes of frame B. Bisector OP of the angle $ct_B$-O-$x_B$ is the primordial photon's world-line. $A_1(t)$-$A_2(t)$ is the rod A at a moment $t$ by the reference clock A; $B_1(t)$-$B_2(t)$ represents rod B at the same moment *by the reference clock B*; these readings of the two reference clocks are *not* simultaneous in either A or B (compare the corresponding simultaneity lines N-O′($t$) and M-O′($t$)). In the case considered here, the proper distance between the events $B_1(t)$ and $B_2(t)$ is the proper length of rod B at the moment $t$ by the B-clock; it is less than the proper length $A_1(t)$-$A_2(t)$ of rod A roughly by a factor of 1.2. However, the interval $B_1(t')$-$B_2(t')$ represents the rod B at a later moment $t' \approx 1.2t$ of B-time, for which its proper length is equal to $l_A^0(t)$.